# Ultrafast Hole Transport and Electron Localization in Superatomic $Re_6Se_8Cl_2$

Jocelyn L. Mendes-Antaya[1], Ye-Jin Kim[1,2], Jonathan M. Michelsen[1], Jessica Yu[3], Shoushou He[3], Xavier Roy[3], Hanzhe Liu[4]*, Scott K. Cushing[1]*

[1] Division of Chemistry and Chemical Engineering, California Institute of Technology, Pasadena, California 91125, United States

[2] Department of Chemistry, Seoul National University, 1 Gwanak-ro, Gwanak-gu, Seoul 08829, Korea

[3] Department of Chemistry, Columbia University, New York, New York 10027, United States

[4] Department of Chemistry, Purdue University, West Lafayette, Indiana 47907, United States

*Correspondence and requests for materials should be addressed to H. L. and S.K.C.

(email: hanzhe@purdue.edu, scushing@caltech.edu)

**Abstract**

Van der Waals superatomic crystals represent an intriguing class of materials, exhibiting unique ultrafast electronic and structural responses arising from their hierarchical crystal structure. These materials bridge the properties of semiconductors, molecular solids, and nanocrystal arrays, hosting a range of many-body interactions that can be controlled with atomically precise structures. To establish the structure-function correlation, it is necessary to understand element-specific electronic states arising from complex superatomic unit cells. Here, we use transient XUV spectroscopy to individually probe the dynamics of photoexcited electrons and holes in $Re_6Se_8Cl_2$ superatomic crystals. We simultaneously measure photodynamics at the Re $O_2$ and Se $M_{4,5}$ edges, enabling decomposition of excited-state dynamics into the contributions of electron, hole, band gap, and phonon dynamics occurring around Re and Se atoms. By correlating spectral features at the Re and Se edges, we infer the existence of flat conduction bands and identify hot hole transfer from Re to Se atoms. We observe phonon oscillations driven by hole-mediated displacive excitations. The measurements highlight that the photodynamics are dominated by charge transfer between metallic sites which drives phonon coupling in the nonequilibrium state. This work provides insights into the optoelectronic applications of superatomic crystals for future tuning of emergent material properties.

## Introduction

The material properties and functions in hierarchical structures are an emerging field in materials science.[1–3] While structure-function relationships are well understood in many cases, they are challenging for materials with complex chemical compositions and hierarchical structures due to the correlated behavior of electrons and phonons across different length scales.[4] Two-dimensional van der Waals (vdW) superatomic crystals are emerging materials that have this challenge. Unlike conventional crystals, superatomic crystals use atomically precise clusters as building blocks that form long-range crystalline order. The materials are composed of discrete, modular clusters that mirror traditional atomic solids.[1,2,5,6] The macroscopic properties of superatomic crystals can be widely tuned by modifications of the chemical composition [7–9] or atomic coordination of cluster building blocks,[10] leading to complex many-body interactions and rich dynamics.

Meanwhile, superatomic crystals have shown promising applications in optoelectronics,[11] thermoelectricity,[12–14] and catalysis.[15] For example, superatomic materials have demonstrated significant thermoelectric performance without systematic material optimization.[16] Recent optical studies reveal efficient quasi-ballistic exciton transport in superatomic crystals up to room temperature, with exciton mobility exceeding Si, making them promising candidates for efficient optoelectronic devices.[17] Other optical studies have provided insight into charge density waves, coherent exciton transport, carrier, and phonon dynamics in $Re_6Se_8Cl_2$.[6,17–19] Further, tuning of electron and phonon bandwidths[1] and superconductivity[9,20,21] have been demonstrated.

The observed material properties are largely due to the intrinsically strong electron–phonon interactions in superatomic crystals.[22,23] Superatomic crystals have a complex unit cell and hierarchical lattice structure, enabling these unique phonon dynamics. Typical phonon spectra consist of both localized molecular vibrations and long-range coherent phonons expected for crystals.[24,25] Meanwhile, electronic excitations create molecular-like excitons that exhibit large binding energies. Despite its fundamental importance, understanding excited-state carrier–phonon dynamics in superatomic crystals is challenging due to the interplay between the complex unit-cell structure, electronic band structures, and various vibrational modes.

In this work, we measure the carrier- and element-resolved ultrafast electronic dynamics in the vdW superatomic semiconductor $Re_6Se_8Cl_2$ (**Figure 1a**), using transient extreme ultraviolet (XUV)

reflection–absorption spectroscopy. In our measurements, carrier- and element-specific relaxation processes following photoexcitation are separately resolved. We measure the thermalization and charge redistribution of holes from Re to Se sites and confirm the existence of flat conduction bands (CBs) through minimal electron thermalization dynamics. Hole-mediated coherent cluster-tilt displacive phonon motions are observed in the transient spectra, supporting carrier-phonon coupling within the two-dimensional layers of $Re_6Se_8Cl_2$. The element-specific observation of flat CBs and the displacive excitation of coherent phonon motions provides insights into the nature of enhanced thermal transport properties within this class of superatomic crystals.[17,25,26] Our observations provide a guide for designing chemical structures and atomically-precise spatial layouts in these hierarchical superatomic crystals, which can potentially enhance carrier transport from the nanoscale to mesoscopic levels.

## Results and Discussion

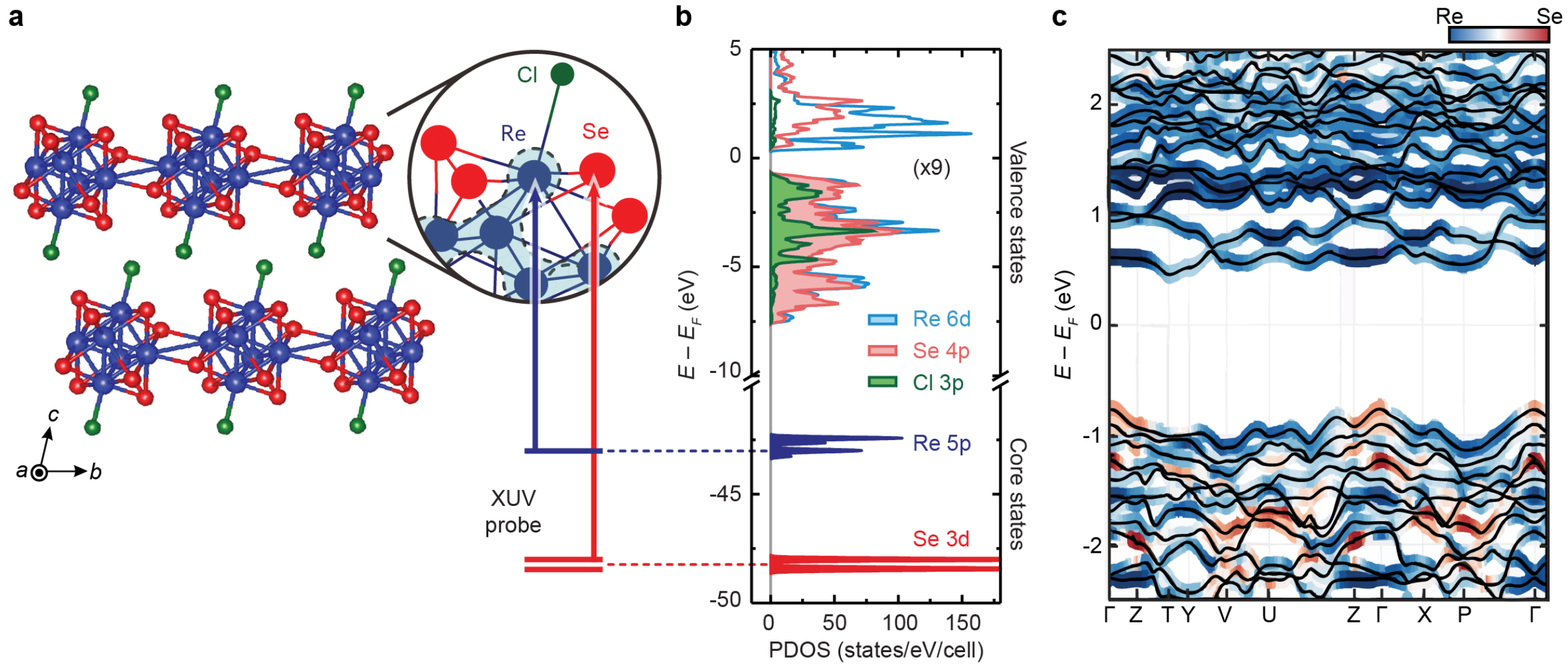

**Figure 1. $Re_6Se_8Cl_2$ crystal and electronic structure. a,** Crystal structure of the $Re_6Se_8Cl_2$ single crystal projected along the *a*-axis. The broadband XUV probe can selectively measure the transient dynamics of charge carriers at both the Re $O_2$ and Se $M_{4,5}$ edges following photoexcitation. **b,** Calculated projected electronic density of states (PDOS). The valence states are composed of Re 6*d*, Se 4*p*, and Cl 3*p* orbitals. Core electrons in Re 5*p* and Se 3*d* orbitals are excited to these states by the XUV probe pulses. The calculation includes the spin-orbit splitting of the core orbitals. $E_F$ is Fermi energy and valence states are scaled by a factor of 9 for better visibility. **c,** Calculated atom-projected band structure of $Re_6Se_8Cl_2$ along the high-symmetry *k*-path. The atomic contributions to the electronic structure indicated by a color gradient from red to blue, representing Se to Re, respectively. The CBs are predominantly composed of Re 6*d* orbitals, while the VBs are hybridized with Re 6*d* and Se 4*p* orbitals.

The layered structure of $Re_6Se_8Cl_2$ is shown in **Figure 1a**. Each layer in this superatomic semiconductor is composed of covalently bonded $Re_6Se_8Cl_2$ clusters arranged in a pseudo-square lattice within the *a*–*b* plane. These two-dimensional layers are stacked along the *c*–axis through vdW interactions, giving rise to a hierarchical crystal structure.[5] This hierarchical structure leads to unusual electron-phonon coupling,[17,18] primarily due to the interplay between bound excitons, localized molecular dynamics, and the long-range collective response. The calculated projected density of states (PDOS) using density functional theory (DFT) predicts an indirect band gap ($E_g$) of 1.2 eV as shown in **Figure 1b**, similar to measured and reported values (see the **Supplementary Information** for details of DFT calculations).[5] The upper valence bands (VBs) have strong hybridization between the Re 6*d* and Se 4*p* orbitals, with minor contributions from Cl 3*p* orbitals. The atomic orbital-resolved electronic band structure, shown in **Figure 1c**, shows that $Re_6Se_8Cl_2$ has an indirect band gap between the valence band maximum (VBM) at the Γ point and the conduction band minimum (CBM) at the T point.[24]

Transient XUV reflection-absorption spectroscopy measures element-specific core-to-valence transitions and can separate electron and hole dynamics at individual elemental X-ray edges.[27] In our experiment, a 3.1 eV (400 nm) excitation pulse generates electron-hole pairs across the band gap of single crystalline $Re_6Se_8Cl_2$. The excitation fluence used in transient XUV experiments was ~9 mJ/cm$^2$, resulting in an excited carrier density of ~$4.1 \times 10^{21}$ cm$^{-3}$. The resulting carrier dynamics are probed by a broadband XUV pulse in a 10° grazing incidence reflection geometry. The XUV pulse simultaneously measures the Re $O_2$ and Se $M_{4,5}$ edges, revealing the local chemical environment around Re and Se sites, as shown in **Figures 2** and **3**, respectively. More details of the experimental configuration can be found in the **Supplementary Information**.

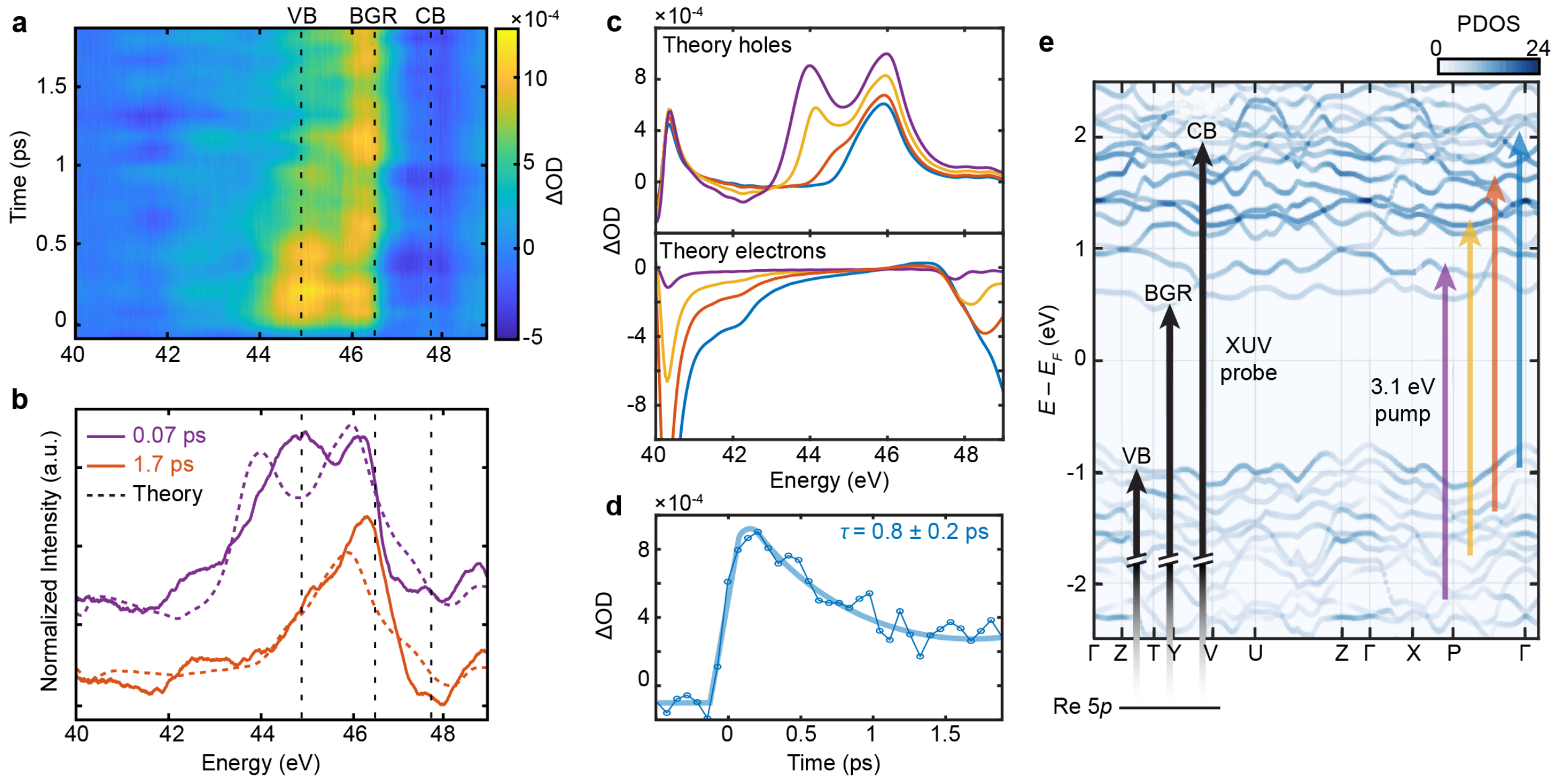

**Figure 2. Carrier relaxation dynamics at the Re $O_2$ edge. a,** Transient XUV reflection-absorption spectra at the Re $O_2$ edge. Three main spectral features for the valence bands (VB), band gap renormalization (BGR), and conduction bands (CB) are denoted as dotted lines. **b,** Experimental and BSE theory calculated transient differential absorption spectra, shown as solid and dotted lines, respectively, at two time delays (0.07 and 1.7 ps) following photoexcitation. The BSE calculated spectra model an initial charge transfer state with electrons filled to the same energy (2.0 eV) above the VBM in both time points. In comparison, holes are filled up to 3.1 eV below the CBM (0.07 ps, purple dashed line) and 1.5 eV below the CBM (1.7 ps, orange dashed line), suggesting strong hole thermalization in the VBs with very minimal electron thermalization. **c,** BSE theory calculated Re $O_2$ edge spectra for holes only (top) and electrons only (bottom). The BSE modeling of holes and electrons uses a state-filling model in which holes are modeled from the VBM down to energies as low as 3.1 eV below the CBM, to consider the possible allowed transitions following the 3.1 eV pump. Similarly, electrons are modeled starting at the CBM and up to 3.1 eV above the VBM. The modeled hole energies in **c** correspond to hole cutoff energies of 3.1 eV (purple), 2.5 eV (yellow), 2.0 eV (orange), and 1.5 eV (blue). The modeled electron energies correspond to electron cutoff energies of 3.1 eV (blue), 2.5 eV (orange), 2.0 eV (yellow), and 1.5 eV (purple). **d,** Kinetic trace of hot hole thermalization at the Re $O_2$ edge averaged across ~44–45 eV. The solid line indicates a fitted curve. **e,** Re projected band structure of $Re_6Se_8Cl_2$. The colored arrows (right) depict the corresponding state-filling electron and hole energies that are modeled in **c** and are color coded accordingly.

We first focus on understanding the photodynamics occurring at the Re sites. **Figure 2a** shows the transient XUV reflection-absorption at the Re $O_2$ edge. The reflection-absorption is defined as $\Delta OD = -\log_{10}(\frac{I_{\text{pump on}}}{I_{\text{pump off}}})$. In this definition, a positive signal (**Fig. 2a**, yellow) represents a transient decrease in XUV reflectivity, which corresponds to an increase in XUV absorption. The two transient spectral lineouts at $\Delta t$ = 0.07 and 1.7 ps show a clear difference between the initial and final excited states (**Fig. 2b**). The positive signal at ~45 eV shifts to higher energies (blueshift) with concomitant decreases in amplitude. The increase in absorption around ~46.5 eV also blueshifts but with a lesser magnitude than the peak at ~45 eV. There is a decrease in absorption (**Fig. 2a**, blue) that is centered around ~47.8 eV and does not shift spectrally or experience a decrease in amplitude.

To understand the origin of the spectral features at the Re $O_2$ and Se $M_{4,5}$ edges of $Re_6Se_8Cl_2$, we use a density functional theory (DFT) and Bethe–Salpeter equation (BSE) approach to model the excited state X-ray edge dynamics. The method uses Quantum ESPRESSO[28,29] and an excited state modification to the OCEAN (Obtaining Core Excitations from the electronic structure and the NIST BSE solver) code.[30–33] The theoretical methods applied here are discussed in more detail in **Supplementary Note S4**. Briefly, to model carrier thermalization, we define two state-filling cutoff energies, $E_h$ and $E_e$. Where $E_h$ is the maximum hole depth in the VBs referenced from the CMB to encompass all possible transitions in momentum space by the pump and $E_e$ is the maximum electron height in the CBs referenced from the VBM. In the case of our 3.1 eV excitation employed in the transient XUV experiments, this means that $E_e$, $E_h \leq 3.1$ eV, and smaller values correspond to carrier thermalization toward the respective band edges of the carriers. $E_h$ and $E_e$ are phenomenological state-filling cutoff energies that parameterize the modeled spectral sensitivity to hole and electron occupations. While these cutoffs describe the bounds of the pump-accessible excitations, they do not imply that a single photoexcited electron–hole pair simultaneously occupies both cutoffs. Additionally, the BSE state-filling model considers dynamically blocking (electrons) and allowing (holes) core-level transitions and does not consider carrier density. More details on these cutoff energies can be found in **Supplementary Note S4.B**.

**Figure 2b** shows experimental and BSE modeled lineouts for two different time delays in the transient spectra. We model the state immediately following photoexcitation (**Fig. 2b**, purple dashed line) based upon an electron cutoff energy of 2.0 eV and a hole cutoff energy of 3.1 eV. This suggests that the increased absorption between ~44–46.5 eV is primarily due to the photoexcited hole population in the VBs. The increased absorption is due to this state-filling effect, where photoinjected holes increase the available density of states in the VBs, leading to enhanced XUV absorption.[34] Similarly, the decrease in absorption from ~46.5–49 eV is well described by photoexcited electrons occupying CBs and blocking XUV transitions, leading to decreased XUV absorption. This theoretical assignment agrees well with experimental spectra immediately following photoexcitation, suggesting that transitions are being probed from deep VBs into CBs near the CBM.

The orange dashed line in **Figure 2b** well describes the transient spectra at later times (1.7 ps). This lineout was modeled with a state-blocking $E_e$ and $E_h$ of 2.0 eV. This suggests that there is significant thermalization of holes at the Re $O_2$ edge following photoexcitation, but minimal thermalization of electrons. Further, the blueshift observed experimentally at ~45 eV in **Figure 2a** is well described by BSE theory modeling hole thermalization at the Re $O_2$ edge, with no electron thermalization modeled. The spectral shift associated with hole thermalization in the VBs occurs on a timescale of 0.8 ± 0.2 ps, with an energy shift of ~0.4 eV (**Fig. 2d**).

The photoinduced changes at the Re $O_2$ edge can be understood as a mixed contribution from photoexcited electrons, holes, and band gap dynamics. To quantify their relative contribution, we apply a singular value decomposition (SVD), described in more detail in **Supplementary Note S3**. The first component is best described by hole thermalization in the VBs, as demonstrated by our BSE theory in **Figure 2c (top)**. The second and third components are well described by band gap renormalization (BGR) and absorption by electrons in the CBs. In addition to photoexcited electron and hole dynamics, SVD reveals that BGR is observed around ~46.5 eV. Microscopically, photoexcitation modifies the Coulomb screening, reducing the band gap.[35] The reduction of the minimum band gap enables XUV transition into the otherwise forbidden gap region, causing a red shift of the XUV transition to the CB immediately following photoexcitation.

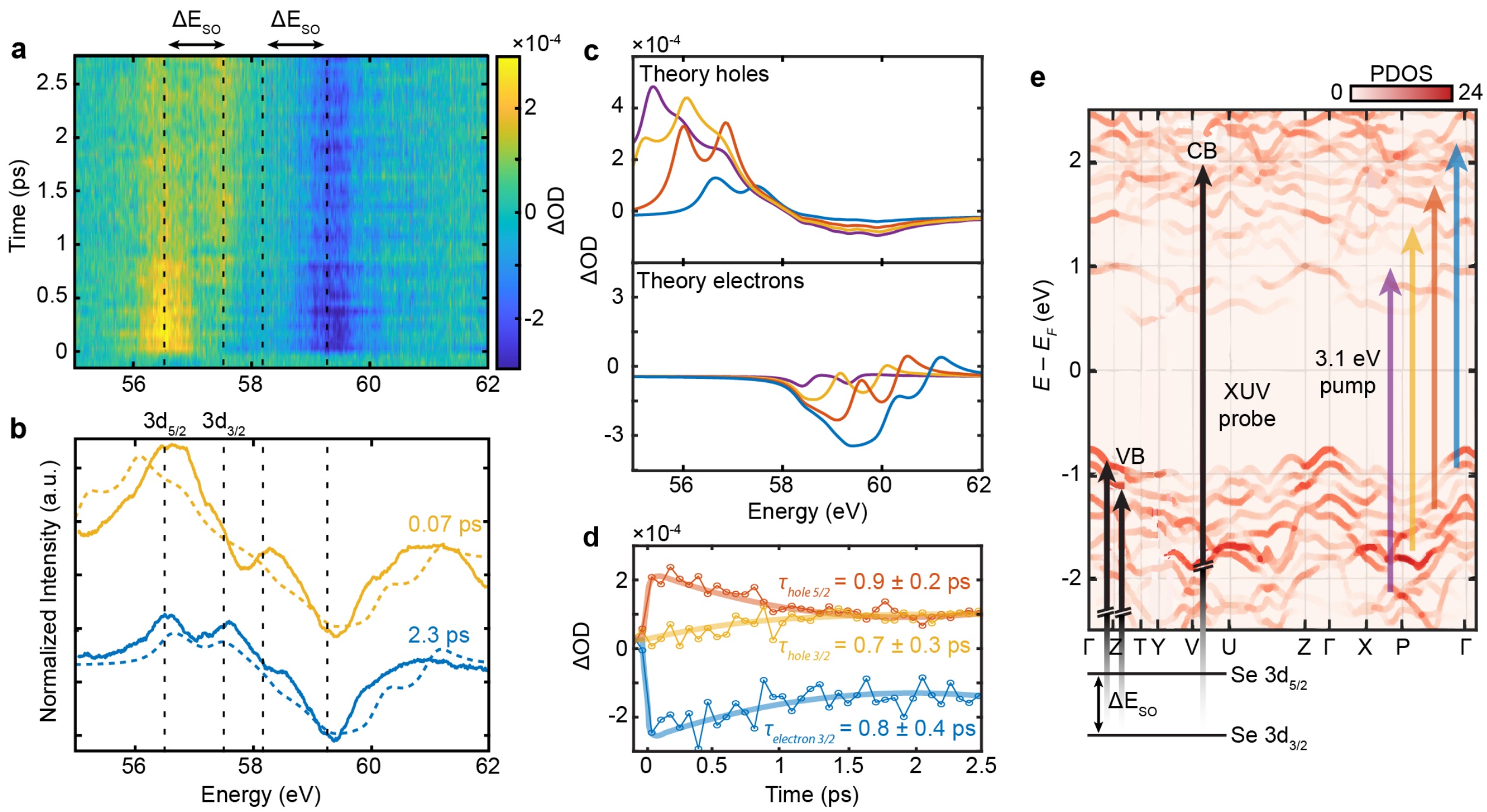

**Figure 3. Carrier relaxation dynamics in Se. a,** Transient XUV reflection-absorption spectra at the Se $M_{4,5}$ edge. **b,** Experimental and BSE calculated transient differential absorption spectra, shown as solid and dashed lines, respectively, at two time delays ($\Delta t$ = 0.07 and 2.3 ps) following photoexcitation. The Se $3d_{5/2}$ and $3d_{3/2}$ transitions are depicted as vertical dashed black lines, with one pair denoting transitions to the VBs and the other pair denoting CB transitions. **c,** As in **Figure 2c**, BSE theory calculated spectra

for holes (top) and electrons (bottom) uses a state-filling model in which $E_h \leq 3.1$ eV, and $E_e \leq 3.1$ eV. The modeled holes correspond to holes filling 3.1 eV (purple), 2.5 eV (yellow), 2.0 eV (orange), and 1.5 eV (blue) below the CBM. The modeled electrons correspond to electrons filling to 3.1 eV (blue), 2.5 eV (orange), 2.0 eV (yellow), and 1.5 eV (purple) above the VBM. **d,** Kinetic profiles for thermalization of holes and electrons at Se sites. The feature at ~56.5 eV attributed to a Se $3d_{5/2}$ transition to the VB has a time constant of 0.9 ± 0.2 ps (red). The feature at ~57.5 eV associated with a Se $3d_{3/2}$ transition to the VB increases in intensity with a time constant of 0.7 ± 0.3 ps (yellow). The electrons in the CB at ~59.3 eV have a thermalization time constant of 0.8 ± 0.4. **e,** Atom-projected band structure for Se atoms where the color scale of the band structure indicates the PDOS of Se and the colored arrows (right) depict the corresponding state-filling electron and hole energies that are modeled in **c**.

We performed similar transient XUV reflectance measurements at the Se $M_{4,5}$ edges around ~57 eV (**Fig. 3a**) and find that the spectral response at the Se $M_{4,5}$ edge is qualitatively different from the Re $O_2$ edge. Immediately after photoexcitation, there is a transient increase in XUV absorption around ~56.5 eV, and two transient decreases in absorption are present at ~58.1 eV and ~59.3 eV, respectively (**Fig. 3b**, yellow solid line). At longer pump-probe time delays, the reduced absorption at ~58.1 eV disappears, and an increased absorption feature appears at ~57.5 eV (**Fig. 3b**, blue solid line).

The transient XUV spectra at the Se $M_{4,5}$ edges reflect local carrier densities around Se atoms and can be understood through photoexcited state-filling effects.[33,34] The Se $M_{4,5}$ edge consists of core-to-valence transitions from two spin-orbit separated Se 3 core levels, $3d_{3/2}$ and $3d_{5/2}$, with ~ 1 eV energy spacing.[36] We assign the transient reflectance changes at ~57.5 eV and ~59.3 eV to an XUV transition from the Se $3d_{3/2}$ core-level to the VBs and CBs, respectively. Similarly, spectral changes at ~56.5 eV and ~58.1 eV are assigned to XUV transitions from the Se $3d_{5/2}$ core level. The ~1 eV spacing of the spin-orbit transitions is consistent with our measured and BSE calculated ground-state core-level absorption spectra shown in **Figure S3** in the **Supplementary Information**.

We further deconvolve the transient dynamics of $Re_6Se_8Cl_2$ at the Se $M_{4,5}$ edges through BSE modeling of photoexcited thermalization dynamics. **Figure 3b** compares BSE modeled transient dynamics and experimental lineouts immediately after photoexcitation (0.07 ps) and following thermalization (2.3 ps). The modeled lineout immediately following photoexcitation is based upon state-filling of electrons up to 3.1 eV above the VBM. In comparison, a hole cutoff energy of 2.5 eV best agrees with the experimental lineout at 0.07 ps (**Fig. 3b**, yellow dashed line). For lineouts following carrier thermalization (**Fig. 3b**, blue dashed line), BSE state-filling of an electron cutoff energy of 3.1 eV and a hole cutoff energy of 1.5 eV best agree with experiment. This suggests that, like carriers at the Re $O_2$ edge, there is little to no electron thermalization, while the holes do thermalize.

The lack of a clear Se $M_{4,5}$ edge thermalization signal and our BSE modeling indicates that electronic thermalization is not readily resolved at the edge. This may arise from electron redistribution within relatively flat CBs of similar Se character that can only be weakly distinguished at the Se edge. Further, this suggests the Se $M_{4,5}$ edge's is sensitive to different CBs than the Re edge, as BSE state-filling of electrons at the Re edge agree better with spectra at lower electron cutoff energies (**Fig. 3c**). We fit the thermalization of holes at the Se $M_5$ edge, an increase in hole signal at the Se $M_4$ edge, and a decrease in electron signal at the $M_4$ edge in **Figure 3d**.

The element-specific dynamics at Re and Se atoms can be further used to infer the distribution of photoexcited carriers in the excited state. As shown in **Figures 2** and **3**, photoexcited electrons are distributed on both Re and Se atoms. Meanwhile, based on the atomic projected band structure (**Fig. 1c**), transient spectra (**Figs. 2a** and **3a**), and BSE calculations (**Figs. 2b** and **3b**), we observe that the hole states have strong Re orbital contributions, particularly near the top of the valence bands. This observation is consistent with our experimental findings that hot holes occur predominantly at the Re sites after photoexcitation and our BSE theory that requires 0.6 eV higher hole cutoff energies on Re sites than Se sites to agree with experiment. As hot holes thermalize at the Re edge, a spectral crossing appears at the Se edge from ~57–58 eV (**Fig. 3a**), where the initial electron spectral weight (blue) after photoexcitation is transferred to holes (yellow) at later times.

This spectral crossing at Se edge could be intuitively understood as a transient increase in hole concentration around the Se atoms during carrier diffusion. Schematically, during the hole thermalization, hot holes migrate from the initial photoexcited states towards the Γ point at the top of the VB, which has increasing contributions from Se atoms (**Fig. 1c**). The net effect of hot hole thermalization is equivalent to hot hole transfer from Re to Se atoms in the unit cell. Given that the spectral crossing timescale at Se sites is comparable to the hole thermalization time at the Re edge (**Figs. 2d** and **3d**), this suggests that the hole transfer from Re to Se atoms slows down overall hole diffusion at Se sites, which leads to the observed spectral crossing at the Se edge.

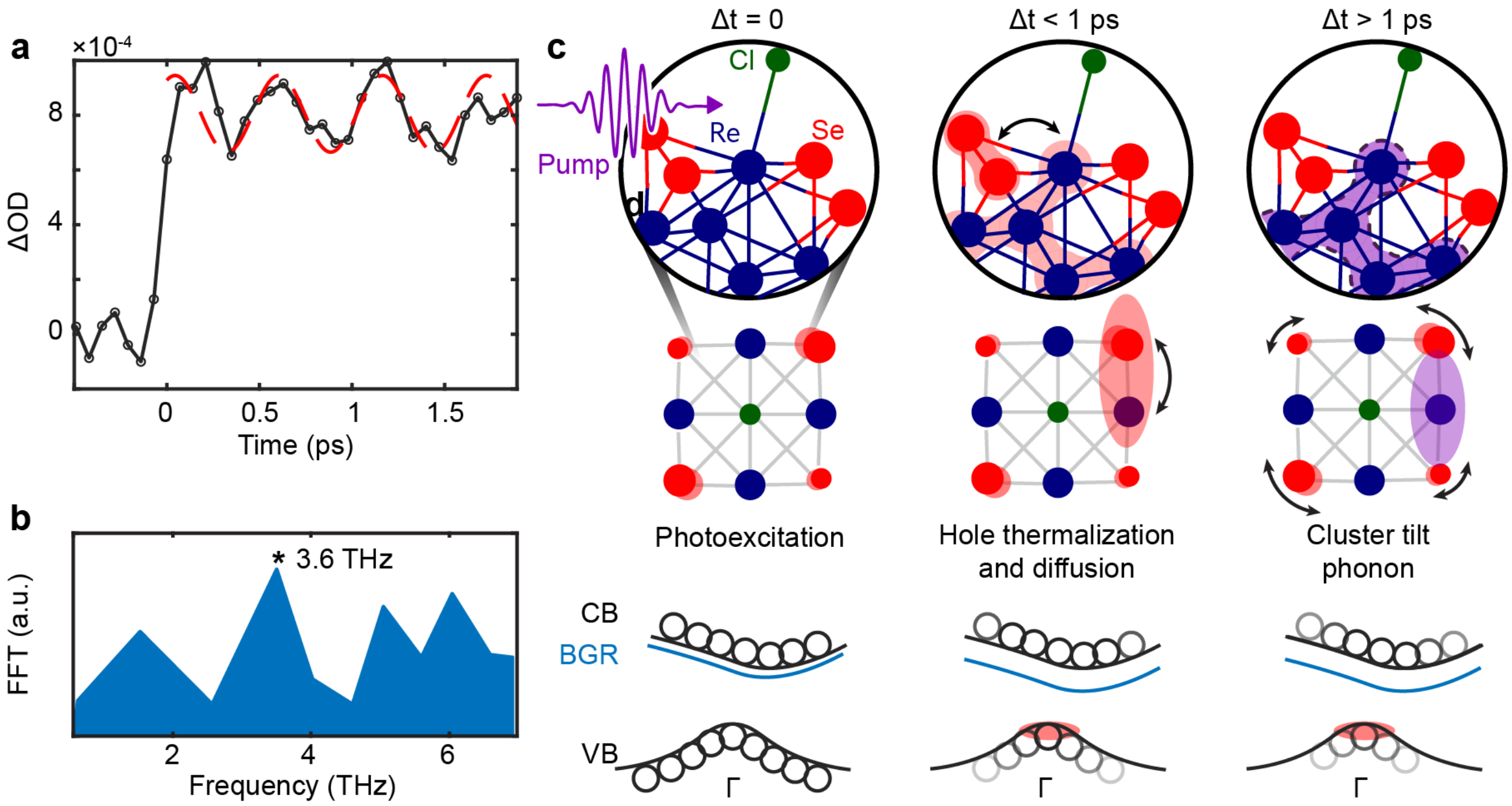

**Figure 4. Ultrafast hole thermalization, diffusion, and phonon oscillations in $Re_6Se_8Cl_2$. a,** Kinetic profile of intensity oscillation averaged across ~45.9 – 46.3 eV, fitted with a sine function (dashed red line). The oscillation period is measured to be 270 fs, which corresponds to 3.6 THz as shown in panel **b**. **b,** Fast Fourier transform (FFT) of the kinetic profile in panel **a**. The highest amplitude oscillation frequency is indicated with an asterisk. **c,** Overall schematics for the photoinduced electronic dynamics in $Re_6Se_8Cl_2$. The atomic view of the time-evolution of the photoexcited states (top), two-dimensional view of the clusters (middle), and charge carrier dynamics depicted onto a simplified band structure (bottom). The element- and carrier-sensitive XUV probe enables elucidation of the initial charge carrier distribution in the VB and CB, and their relaxation dynamics with coupling to displacive phonons. The cluster tilt phonon mode accompanies vibrations of the Se atoms at the corner, leading to the in-plane distortion of a cluster.

We measure phonon oscillations in our transient XUV absorption spectra at the Re $O_2$ edge (**Fig. 2a**). The kinetic profile and a fit curve display a time-dependent modulation of intensity with a periodicity of 270 fs (3.6 THz) (**Fig. 4a**). The periodicity aligns with a coherent phonon motion associated with the twisting of superatomic clusters in an $Re_6Se_8Cl_2$ layer, primarily driven by the displacement of the corner Se atoms (**Figs. 4b** and **4c**).[18] These two-dimensional phonon modes, fall within the frequency range of 1.69–5.25 THz and can be linked to the electronic modulation of free charge carriers. The oscillation has a cosine character at $t_0$ that suggests the mechanism of this oscillation is due to displacive excitation of coherent phonons (DECP) rather than impulsive stimulated Raman scattering (ISRS).[18,37,38] Specifically, this coherent oscillation is only present in the positive absorption signal at the Re $O_2$ edge, suggesting that DECP is hole-mediated.[39] Mechanistically, this can be understood as photoexcited holes transporting through metallic sites from Re to Se atoms in $Re_6Se_8Cl_2$ single crystals and the thermalization and the associated intravalley charge redistribution dynamics across different momenta mediate coherent phonon oscillations, facilitating energy dissipation.

Notably, hole thermalization is observed at both the Re $O_2$ and Se $M_{4,5}$ edges in the more dispersive VBs, while electron thermalization is minimal in the flat CBs. Our consensus is that the electrons experience very little relative thermalization in energy compared to holes in the $Re_6Se_8Cl_2$ superatomic crystal, suggesting that the VBs are much more dispersive, with more delocalized electronic states than the CBs. We further find that hole transport from Re to Se atoms displacively excites coherent cluster twisting phonons, which facilitate energy transfer towards adjacent interclusters within 2D metal layers.

In summary, we report transient XUV reflection–absorption spectroscopy on the photoexcited dynamics in $Re_6Se_8Cl_2$ superatomic crystals. Photoinduced electron, hole, phonon, and band gap dynamics are observed and separated at both Re and Se sites. By correlating the ultrafast dynamics at Re and Se atoms, we infer the existence of flat conduction bands and identify an ultrafast hole transfer from Re to Se atoms. Our results provide a detailed picture of element- and carrier-specific dynamics in superatomic crystals, as well as new insights into the real-space charge transfer processes in $Re_6Se_8Cl_2$, paving the way for a deeper understanding of more complex many-body interactions in superatomic crystals. These observations underscore the impact of nonequilibrium many-body effects and enhance the understanding of photoexcited electron and hole dynamics across different elements in superatomic crystals. These insights are vital for advancing the application of novel photophysics and for driving the development of exotic phases, such as doping-induced superconductivity,[9] in superatomic crystals.

**Methods**

Details about sample preparation and characterization, the transient XUV spectrometer, spectral analysis, and theoretical modeling of excited state spectra using DFT and the BSE can be found in the Supplementary Information.

**Acknowledgements**

Work on superatomic materials at Columbia University is supported by the Air Force Office of Scientific Research (AFOSR) under MURI grant number FA9550-25-1-0288. The computations presented here were conducted in the Resnick High Performance Computing Center, a facility supported by the Resnick Sustainability Institute at the California Institute of Technology. This research used resources of the National Energy Research Scientific Computing Center, a DOE Office of Science User Facility supported by the Office of Science of the U.S. Department of Energy under Contract No. DE-AC02-05CH11231 using NERSC award BES-ERCAP0024109. Y.-J.K. acknowledges support by the Liquid Sunlight Alliance (the U.S. Department of Energy, Office of Science, Office of Basic Energy Sciences, Fuels from Sunlight Hub under award number DE-SC0021266). J.L.M. acknowledges support by the National Science Foundation Graduate Research Fellowship Program under grant no. 1745301. H.L. acknowledges support from the National Science Foundation under award number 2325212.

Supplementary Information

# Ultrafast Hole Transport and Electron Localization in Superatomic $Re_6Se_8Cl_2$

Jocelyn L. Mendes-Antaya[1], Ye-Jin Kim[1,2], Jonathan M. Michelsen[1], Jessica Yu[3], Shoushou He[3], Xavier Roy[3], Hanzhe Liu[4]*, Scott K. Cushing[1]*

[1] Division of Chemistry and Chemical Engineering, California Institute of Technology, Pasadena, California 91125, United States

[2] Department of Chemistry, Seoul National University, 1 Gwanak-ro, Gwanak-gu, Seoul 08829, Korea

[3] Department of Chemistry, Columbia University, New York, New York 10027, United States

[4] Department of Chemistry, Purdue University, West Lafayette, Indiana 47907, United States

*Correspondence and requests for materials should be addressed to H. L. and S.K.C.

(email: hanzhe@purdue.edu, scushing@caltech.edu)

**Supplementary Note 1. $Re_6Se_8Cl_2$ sample preparation**

$Re_6Se_8Cl_2$ crystals were synthesized via chemical vapor transport adapted from previous reports.[1,2] Re (330 mg, 1.77 mmol), Se (190 mg, 2.41 mmol) and $ReCl_5$ (200 mg, 0.55 mmol) were ground and mixed using an agate mortar inside the glove box, and subsequently pressed into a pellet. The pellet was sealed in a 20 cm quartz tube under vacuum and heated in a tube furnace at a rate of 1 °C/min to 1100 °C for 3 days. To achieve large single crystals, the tube was then cooled over six hours to a temperature gradient of 1100–1030 °C and held for 270 hours. The tube was further cooled to room temperature over 12 hours, and the furnace was then shut off. To extract pristine $Re_6Se_8Cl_2$ single crystals, a gradient of 300–25 °C was used to separate any excess volatile components, yielding ~1–3 mm single crystals at the top section of the tube. Typical bulk samples have a few millimeters in lateral dimensions. **Figure S1** shows a picture of a typical sample piece used in the transient XUV experiments.

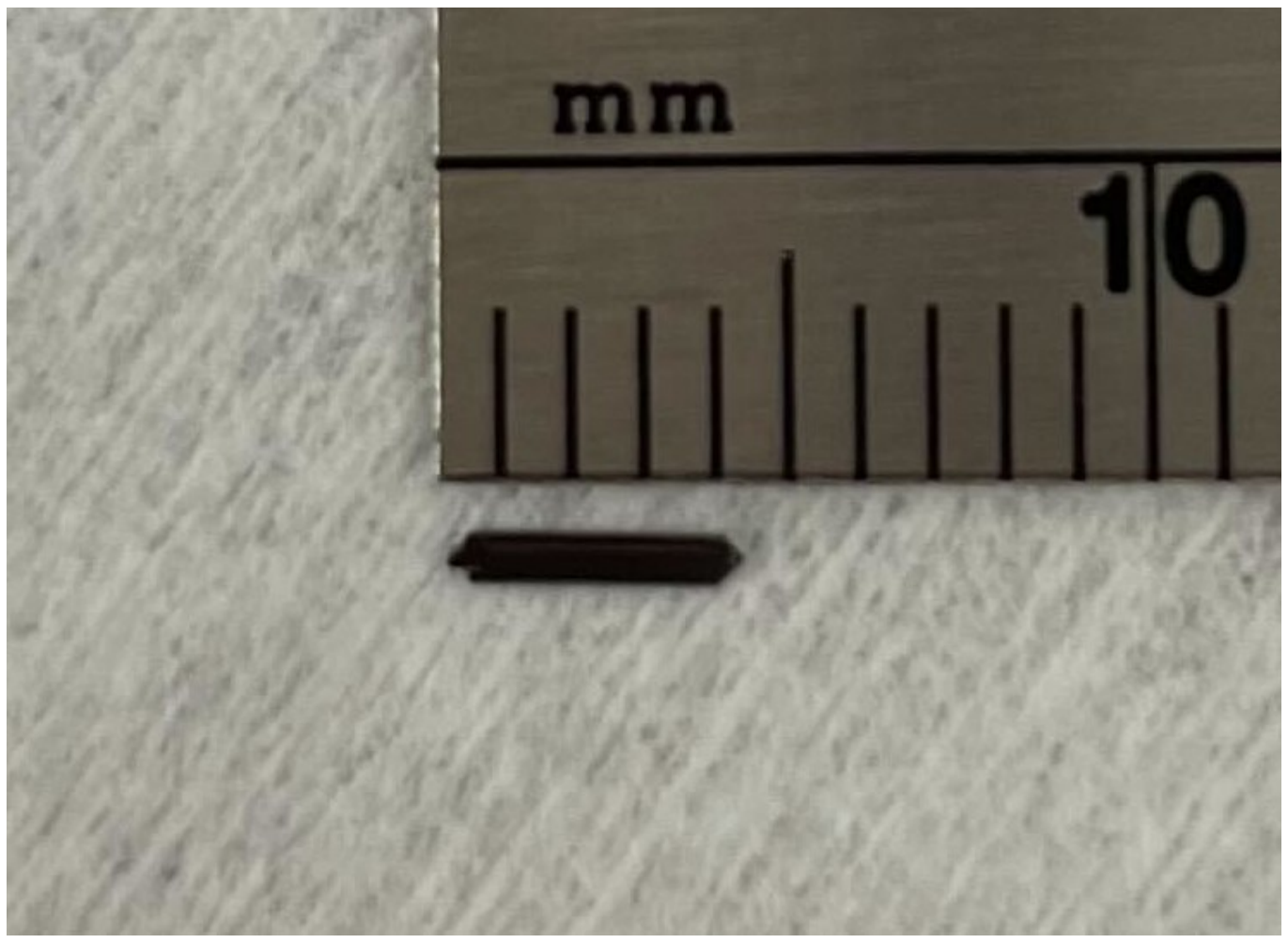

**Figure S1.** A typical mm-size single crystalline bulk $Re_6Se_8Cl_2$ sample used in our transient XUV measurements.

**Supplementary Note 2. Transient extreme ultraviolet spectrometer**

The transient extreme ultraviolet (XUV) reflection-absorption spectrometer is described previously.[3–5] Briefly, a Legend Elite Duo Ti:Sapphire system (Coherent Inc.) with 1 kHz, 35 fs, 13 mJ pulses centered around 800 nm is split by a 75:25 beam splitter. ~3 mJ generates a few-cycle white light beam (<6 fs, 550–950) for high harmonic generation (HHG) and ~10 mJ is used for the pump path. The pump path employs the 400 nm frequency doubled output of a BBO crystal with p-polarization, pumped with the 800 nm output from the Ti:Sapphire laser for an above band gap photoexcitation of the $Re_6Se_8Cl_2$ single crystals. The optical excitation fluence used in these experiments was ~9 $mJ/cm^2$, resulting in an excited carrier density of ~$4.1 \times 10^{21}$ $cm^{-3}$.[6]

Transient reflection is measured at a 10° grazing incidence (80° from normal incidence) geometry and measures the varying delay times between the pump and probe pulses using an optomechanical delay stage. The photoexcited dynamics were probed with an XUV pulse produced through HHG in argon with an s-polarized few-cycle white light pulse driving HHG. The fundamental white light beam is removed with a 100 nm thick Al filter (Luxel). The generated XUV continuum is used to probe the Re $O_2$ and Se $M_{4,5}$ absorption edges between 40–60 eV. An edge-pixel referencing scheme uses signal free spectral regions to denoise intensity fluctuations in the spectra.[7]

## Supplementary Note 3. Spectral analysis

The transient XUV spectra at the Re $O_2$ edge consists of contributions from electrons, holes, and band shifts. We apply singular value decomposition (SVD) to all pump-probe delays within the transient spectrum to extract the weights of relevant dynamics and reconstruct the transient spectrum at the Re $O_2$ edge using the fewest number of singular values. The SVD analysis factorizes the two-dimensional matrix of the transient spectrum by rotating (U), scaling (S), and then again rotating (V) the matrix. The scaling matrix contains the singular values along its diagonal, which we refer to as the number of components. The number of singular values or components in the scaling matrix is selected to minimize the total number of singular values, where at the Re $O_2$ edge three singular values are needed.

Following SVD, we ascribe component 1 to holes, component 2 to electrons, and component 3 to band gap renormalization (BGR). **Figure S2** shows the components of the SVD analysis. Component 1 well describes excited state hole effects, attributed to an increase in absorption at the edge. Due to the minimal band gap renormalization shift, and minimal thermalization in the electron component with respect to time, components 2 and 3 are combined in a weighted average to visualize not only the spectral shift associated with BGR but also minimal electron thermalization. It is worth noting that the lineouts in component 1 are an order of magnitude more intense than those in the weighted average of components 2 and 3. The reconstructed spectra in **Figure S2(c)** is well described by three components, where the band gap renormalization shift is not observed with less than three components.

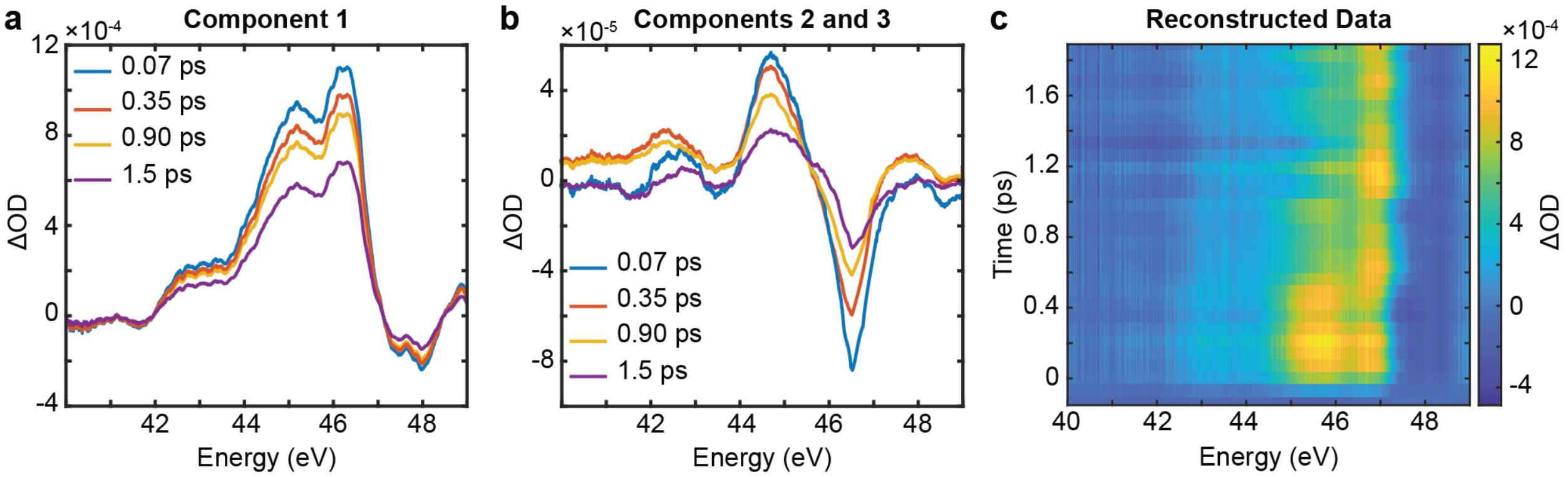

**Figure S2.** Component 1 **(a)** of the SVD analysis at the Re $O_2$ edge. Components 2 and 3 **(b)** are plotted as a weighted average as they are an order of magnitude weaker than component 1. The SVD reconstruction **(c)** agrees with transient XUV spectra of the Re $O_2$ edge shown in **Figure 2a** in the main text.

## Supplementary Note 4. Ab initio modeling

#### 4.A. Density functional theory

Geometry optimization and density functional theory (DFT) calculations were conducted using the Quantum ESPRESSO package with norm-conserving general gradient approximation (GGA), Perdew– Burke–Ernzerhof (PBE) pseudopotentials.[8–10] Geometry optimizations employed an input $Re_6Se_8Cl_2$ unit cell containing 16 atoms and were based upon the lattice parameters $a = b = c =$ 6.67 Å. A 300 Rydberg kinetic energy cutoff and an 8 × 8 × 8 Monkhorst– Pack k-mesh sampling of the first Brillouin zone was used to sample 250 bands. The structures employed in the DFT and BSE calculations described below were fully relaxed and calculations were performed to a convergence of $10^{-8}$ eV/atom with the forces on ions under $10^{-3}$ eV/Å.

#### 4.B. Bethe–Salpeter equation theory of core-level spectra

The fully relaxed parameters used in the DFT calculations discussed in **Supplementary note 4.A** were applied to the BSE calculations of the $Re_6Se_8Cl_2$ ground state reflectivity at the Re $O_2$ and Se $M_{4,5}$ edges. The method uses Quantum ESPRESSO[9,10] and an excited state modification to the OCEAN (Obtaining Core Excitations from the *Ab initio* electronic structure and the NIST BSE solver) code.[11–13] The BSE calculation employed a screening mesh of 2 × 2 × 2, a dielectric constant of 21.8 at 293 K, a 4.0 Bohr screening radius, and a 0.7 scaling factor for the Slater G parameter. A spectral broadening of 0.25 eV was applied for all calculated spectra.

The calculated ground state reflectivity of $Re_6Se_8Cl_2$ is plotted in **Figure S3** and is compared to experimentally measured XUV ground states at either edge. As shown in **Figure S3**, the BSE calculated ground state agrees well with the measured ground state reflectivity of the sample.

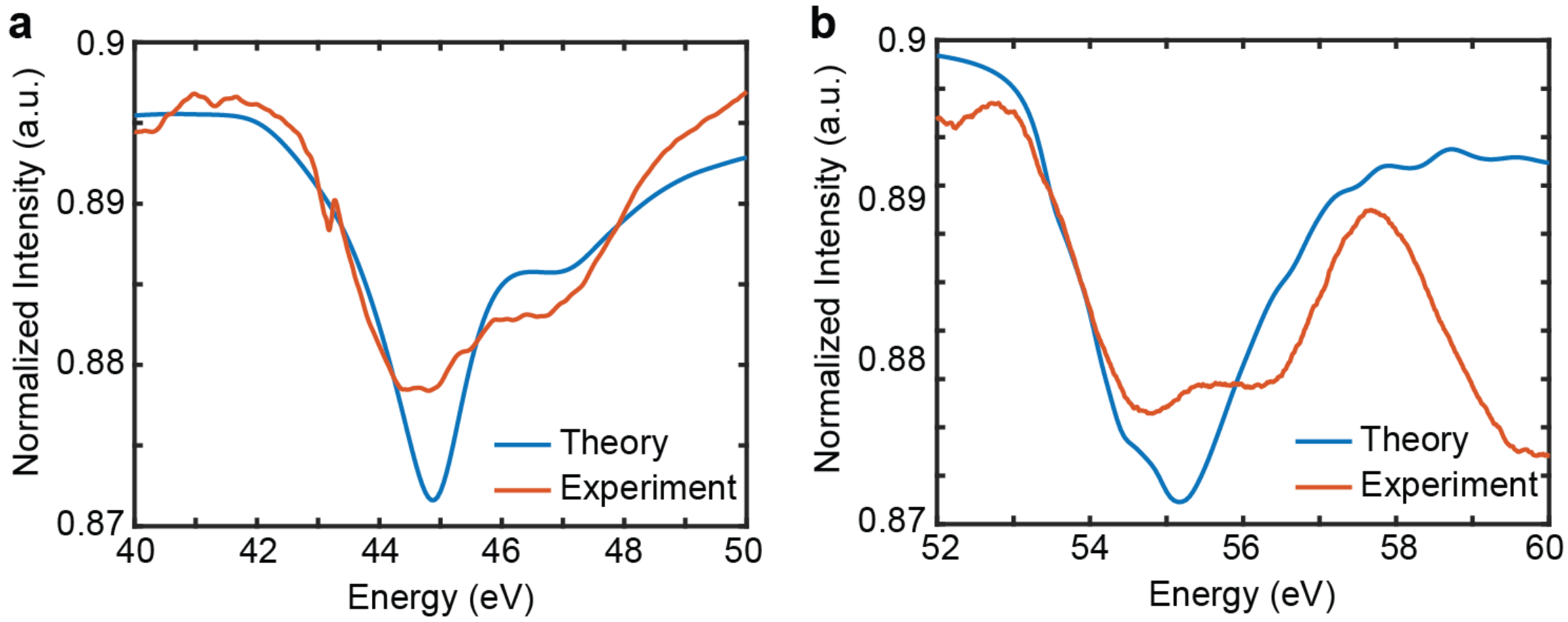

**Figure S3.** Ground state reflectivity at the **(a)** Re $O_2$ and **(b)** the Se $M_{4,5}$ edges. BSE theory of the edges (blue) is compared to experimental static XUV reflectivity (orange) for good agreement.

Following ground state calculations of the XUV reflectivity as the Re $O_2$ and Se $M_{4,5}$ edges, excited state calculations were performed using a modification to the OCEAN code discussed previously.[14] The modification to the OCEAN code enables selective excitation to valence and conduction bands at different points in momentum space, to simulate state-filling of electrons and holes following photoexcitation. To model above band gap photoexcitation, we populate the conduction band with electrons and similarly populate the valence band with holes to model the state-filling of carriers following photoexcitation. This population of electrons and holes effectively forbids or allows XUV transitions to specific points in k-space but does not account for carrier density, therefore, the modeled peak position and *relative* intensity is used to compare experiment to theory. We model our transient XUV spectra by calculating the differential absorption between the calculated ground state and the calculated excited state spectra. **Figures 2** and **3** in the main text demonstrate agreement between BSE modeled excited states at the Re $O_2$ and Se $M_{4,5}$ edges, respectively.

Specifically, to model carrier thermalization and excited-state state filling of electrons and holes, we employ two "cutoff" energies: $E_h$ and $E_e$. Where $E_h$ is the maximum hole depth possible following 3.1 eV excitation the valence bands and $E_e$ is the maximum electron energy accessible following 3.1 eV excitation in the conduction bands. Here, $E_e$ is defined as the energy separation between the CBM and the deepest depopulated valence state, while $E_e$ is defined as the energy separation between the VBM and the highest populated conduction state. **Equations S1** and **S2** depict this relation:

$$E_h = E_{CBM} - E_{occ}^h \quad \textbf{(S1)}$$

$$E_e = E_{occ}^e - E_{VBM} \quad \textbf{(S2)}$$

Here, $E_{occ}^h$ and $E_{occ}^e$ are the energetic positions within the band structure of the electrons and holes that are being applied through the excited-state BSE modeling. Thus, $E_h$ and $E_e$ are employed to parameterize the pump-accessible excited carrier manifold rather than directly visualize carrier energies referenced to their own band edges. For the 3.1 eV pump employed in our XUV experiments, the largest allowed values are $E_e$, $E_h \leq 3.1$ eV, and smaller values correspond to carrier thermalization toward the respective band edges of the carriers.

This parameterization allows us to separably map the thermalization of electrons and holes. **Figures S4** and **S5** show different energy holes **(a)** and electrons **(b)** to model thermalization effects immediately following excitation, all the way to thermalization at the band edge at the Re $O_2$ and Se $M_{4,5}$ edges, respectively. **Figure S4(c)** presents the combination of spectra calculated in **S4(a)** and **S4(b)** where a fixed electron cutoff energy of 2.0 eV is applied to mimic little-to-no electron thermalization at the Re $O_2$ edge, and a varied hole cutoff energy is employed to model ultrafast hole cooling to the valence band edge. This scenario agrees best with experimental XUV spectra at the Re $O_2$ edge. Similarly, **Figure S5(c)** presents the combination of spectra calculated in **S5(a)** and **S5(b)** at the Se $M_{4,5}$ edge, however, a fixed electron cutoff energy of 3.1 eV is applied to mimic little-to-no electron thermalization, and a varied hole cutoff energy is employed to model ultrafast hole cooling to the valence band edge. This scenario agrees best with experimental spectra at the Se $M_{4,5}$ edge and suggests that higher energy electrons are being probed at the Se $M_{4,5}$ edge due to the element-specific nature of the core-level probe. Different projected densities of states are probed by different core-valence transitions. However, in both the case of Re and Se atoms, very minimal electron thermalization is observed spectrally, as supported by BSE theory.

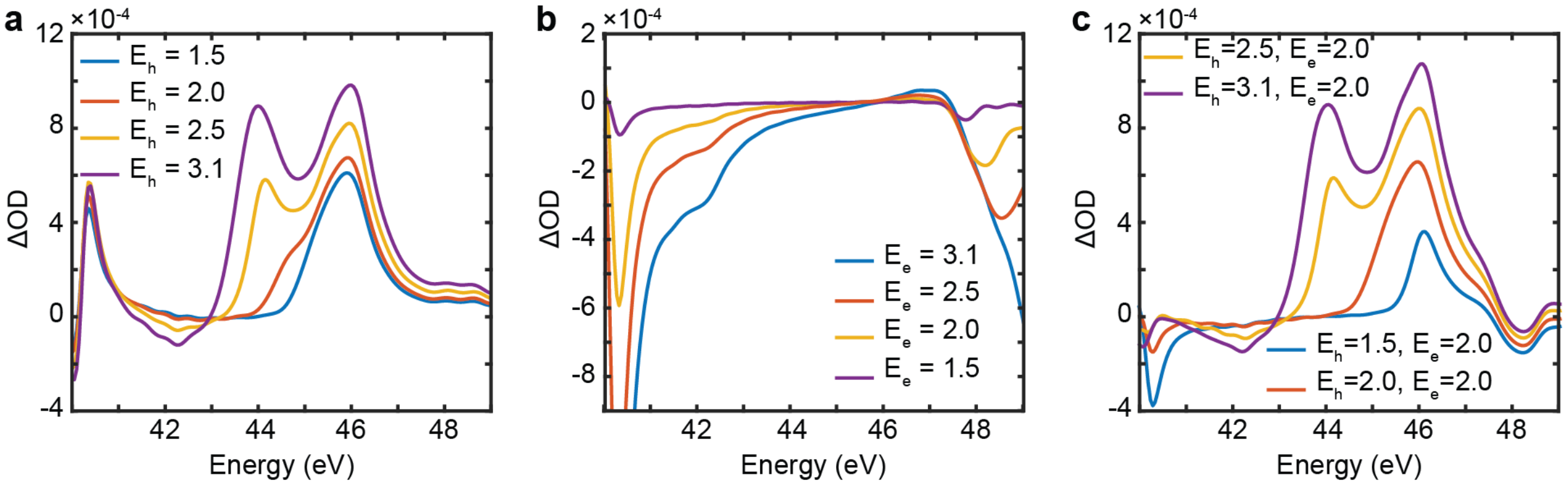

**Figure S4.** BSE modeled excited-state carrier dynamics for holes **(a)**, electrons **(b)**, and holes and electrons together **(c)** at the Re $O_2$ edge. The electron and hole cutoff energies correspond to the VBM and CBM of the $Re_6Se_8Cl_2$ electronic structure respectively.

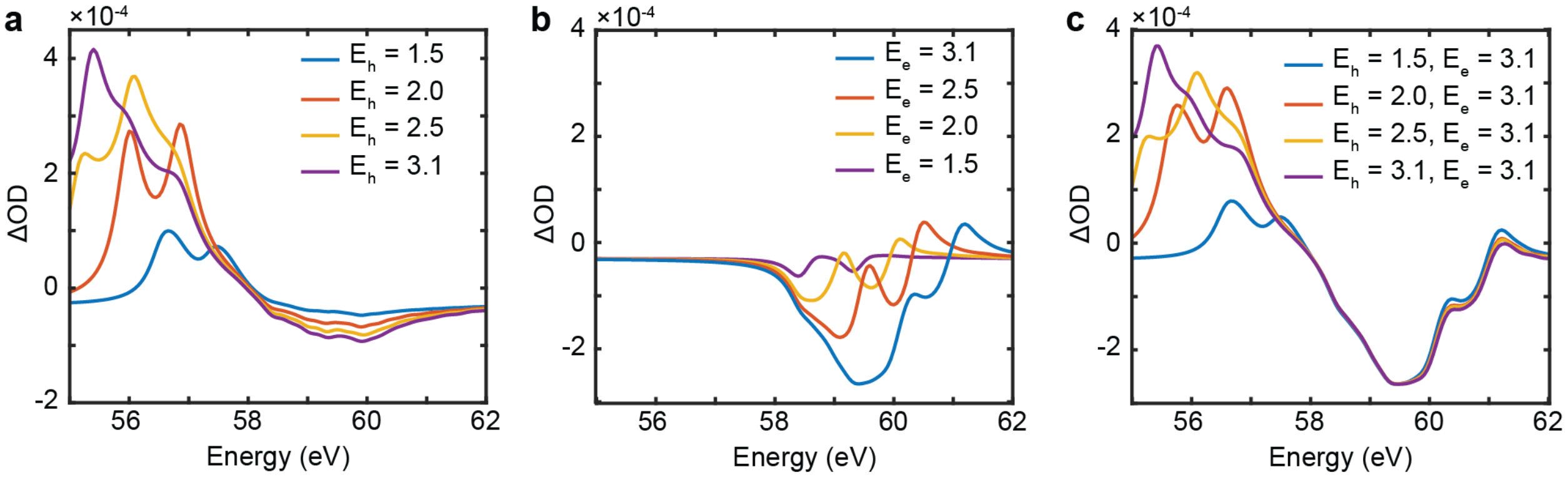

**Figure S5.** BSE modeled excited-state carrier dynamics for holes **(a)**, electrons **(b)**, and holes and electrons together **(c)** at the Se $M_{4,5}$ edge. A higher electron energy is needed in **(c)** to agree with experiment compared to the Re $O_2$ edge, suggesting higher energy electrons are selectively probed at the Se $M_{4,5}$ edge due to the element-sensitivity of the transient XUV technique.